
%
%
%
%
%
\input harvmac
%
%

%
%

\def\Title#1#2{\ifx\answ\bigans \nopagenumbers
\abstractfont\hsize=\hstitle\rightline{#1}%
\vskip 1in\centerline{\titlefont #2}\abstractfont\vskip .5in\pageno=0
\else \rightline{#1}
\vskip .8in\centerline{\titlefont #2}
\vskip .5in\pageno=1\fi}
\ifx\answ\bigans

\else

 \font\absi=cmmi10 scaled\magstep1
\font\absis=cmmi7 scaled\magstep1 \font\absiss=cmmi5 scaled\magstep1
\font\abssy=cmsy10 scaled\magstep1 \font\abssys=cmsy7 scaled\magstep1
\font\abssyss=cmsy5 scaled\magstep1 
\skewchar\absi='177 \skewchar\absis='177 \skewchar\absiss='177
\skewchar\abssy='60 \skewchar\abssys='60 \skewchar\abssyss='60
\fi
%
%
\def\overleftrightarrow#1{\buildrel\leftrightarrow\over#1}

\def\eg{{\it e.g.}}

\def\etc{{\it etc.}}
\def\ket#1{\vert#1\rangle}
\def\bra#1{\langle#1\vert}
\def\phit{{\tilde \phi}_0}

\def\psih{{\tilde \psi}}
\font\ticp=cmcsc10

\def\ajou#1&#2(#3){\ \sl#1\bf#2\rm(19#3)}
\def\apm{{\alpha^\prime}}

\def\Hsl{{\,\raise.15ex\hbox{/}\mkern-12mu H}}
\def\Dsl{\,\raise.15ex\hbox{/}\mkern-13.5mu D}
\def\Ssl{{\,\raise.15ex\hbox{/}\mkern-10.5mu S}}

%
%
\lref\BDDL{T. Banks, A. Dabholkar, M.R. Douglas, and M O'Loughlin, ``Are
horned particles the climax of Hawking evaporation?'' Rutgers preprint
RU-91-54.}
\lref\RST{J.G. Russo, L. Susskind, and L. Thorlacius, ``Black hole
evaporation in 1+1 dimensions,'' Stanford preprint SU-ITP-92-4.}
\lref\HoWi{C.F.E. Holzhey and F. Wilczek, ``Black holes as elementary
particles,'' IAS preprint IASSNS-HEP-91/71.}
\lref\DuLu{M.J. Duff and J.X. Lu, ``Elementary five-brane solutions of D=10
supergravity,''\ajou Nucl. Phys. &B354 (91) 141.}
\lref\WittSton{E.~Witten, ``On black holes and string theory,'' IAS
preprint (Lecture
notes from Strings and Symmetries '91, Stony Brook).}
\lref\alf{M. Alford and A. Strominger, to appear.}
\lref\evan{C. Callan, S.B. Giddings, J. Harvey and A. Strominger,
``Evanescent Black Holes'' preprint UCSB-TH-91-54, hepth@xxx/9111056
(1991), to appear in {\sl Phys. Rev.} {\bf D}.}
\lref\AW{J. J. Atick and E. Witten, ``The Hagedorn transition and the number
of degrees of freedom of string theory'',\ajou Nucl. Phys. & B310 (88) 291;
W. Fischler and J. Polchinski, unpublished.}
\lref\BSW{M. Bowick, L. Smolin and R. Wijewardhana, Phys. Rev. Lett. ???}
\lref\PSSTW{J. Preskill, ``Quantum hair'', CALT-68-1671 (1990);
J. Preskill, P. Schwarz, A. Shapere, S. Trivedi and
F. Wilczek, ``Limitations on the statistical description of black holes,''
IAS preprint IASSSNS-HEP-91/34.}
\lref\DLP{L.J. Dixon, J. Lykken, and M.E. Peskin, ``N=2 superconformal
symmetry and SO(2,1) current algebra,''\ajou Nucl. Phys. & B325 (89) 329.}
\lref\Gibb{G.W. Gibbons, ``Antigravitating black hole solitons with scalar
hair in N=4 supergravity,''\ajou Nucl. Phys. &B207 (82) 337\semi
G.W. Gibbons and K. Maeda, ``Black holes and membranes in
higher-dimensional theories with dilaton fields,''\ajou Nucl. Phys. &B298
(88) 741}
\lref\ddd{A. Shapere, S. Trivedi and F. Wilczek, ``Dual dilaton dyons,''
IASSNS-HEP-91/33}
\lref\BarsBH{I. Bars, ``String propagation on black holes,'' USC preprint
USC-91/HEP-B3; ``Curved space-time strings and black holes,'' USC preprint
USC-91/HEP-B4.}
\lref\ILS{N. Ishibashi, M. Li, and A.R. Steif, ``Two dimensional charged
black holes in string theory,'' UCSB preprint UCSB-91-28.}
\lref\EFR{S. Elitzur, A. Forge, and E. Rabinovici, ``Some global aspects of
string compactifications,'' Hebrew University preprint RI-143-90.}
\lref\RSS{M. Ro\v cek, K. Schoutens, and A.
Sevrin, ``Off-shell WZW models in extended superspace,'' IAS preprint
IASSNS-HEP-91-14.}
\lref\Witt{E. Witten, ``On string theory and black holes,''\ajou Phys. Rev.
&D44 (91) 314.}
\lref\CHS{C. Callan, J. Harvey and A. Strominger, ``Worldsheet
approach to
heterotic solitons and instantons,''\ajou Nucl. Phys. &B359 (91) 611.}
\lref\HoSt{G. Horowitz and A. Strominger, `` Black strings and
$p$-branes,''\ajou Nucl. Phys. &B360 (91) 197.}
\lref\worm{S. Coleman, ``Black holes as red herrings: Topological
fluctuations and the loss of quantum coherence,''\ajou Nucl. Phys. &B307
(88)
867, S.B. Giddings and A. Strominger, ``Loss of incoherence and
determination of coupling constants in quantum gravity,''\ajou Nucl. Phys.
&B307 (88) 854.}
\lref\BaNe{I. Bars and D. Nemeschansky, ``String propagation in backgrounds
with curved space-time,''\ajou Nucl. Phys. &B348 (91) 89.}
\lref\Bars{I. Bars, ``Heterotic superstring vacua in four dimensions based
on non-compact affine current algebras,''\ajou Nucl. Phys. &B334 (90) 125.}
\lref\MSW{G. Mandal, A Sengupta, and S. Wadia, ``Classical solutions of 2d
string theory,'' IAS preprint IASSNS-HEP-91/10.}
\lref\MaSh{E.J. Martinec and S.L. Shatashvili, ``Black hole physics and
Liouville theory,'' Chicago preprint EFI-91-22.}
\lref\DVV{R. Dijkgraaf, H. Verlinde, and E. Verlinde, ``String propagation
in a
black hole geometry,'' Princeton/IAS preprint PUPT-1252=IASSNS-HEP-91/22.}
\lref\GiSt{S.B. Giddings and A. Strominger, ``Exact black fivebranes in
critical superstring theory,''\ajou Phys. Rev. Lett. &67 (91) 2930.}
\lref\HoHo{J.H. Horne and G.T. Horowitz, ``Exact black string solutions in
three dimensions,'' UCSB preprint UCSBTH-91-39.}
\lref\BPS{T. Banks, M. Peskin and L. Susskind, ``Difficulties for
the evolution of pure states into mixed states,'' \ajou Nucl. Phys.
& B244 (84) 125.}
\lref\AW{J. J. Atick and E. Witten, ``The Hagedorn transition and the number
of degrees of freedom of string theory'',\ajou Nucl. Phys. & B310 (88) 291;
W. Fischler and J. Polchinski, unpublished.}
\lref\BSW{M. Bowick, L. Smolin and R. Wijewardhana, Phys. Rev. Lett. ???}
\lref\preskill{J. Preskill, ``Quantum hair'',\ajou Phys. Scrip. &T36
(91) 259.}
\lref\PresComm{J. Preskill, private communication.}
\lref\PSSTW{J. Preskill, P. Schwarz, A. Shapere, S. Trivedi and
F. Wilczek, ``Limitations on the statistical description of black holes,''
IAS preprint IASSSNS-HEP-91/34.}
\lref\dollar{S. W. Hawking, ``The unpredictability of quantum gravity,''
\ajou Comm. Math. Phys. &87 (83) 395.}
\lref\DLP{L.J. Dixon, J. Lykken, and M.E. Peskin, ``N=2 superconformal
symmetry and SO(2,1) current algebra,''\ajou Nucl. Phys. & B325 (89) 329.}
\lref\GHS{D. Garfinkle, G. Horowitz, and A. Strominger, ``Charged black holes
in string theory,''\ajou Phys. Rev. &D43 (91) 3140.}
\lref\cghs{C. Callan, S.B. Giddings, J. Harvey and A. Strominger, to appear.}
\lref\BarsBH{I. Bars, ``String propagation on black holes,'' USC preprint
USC-91/HEP-B3; ``Curved space-time strings and black holes,'' USC preprint
USC-91/HEP-B4.}
\lref\ILS{N. Ishibashi, M. Li, and A.R. Steif, ``Two dimensional charged
black holes in string theory,'' UCSB preprint UCSB-91-28.}
\lref\EFR{S. Elitzur, A. Forge, and E. Rabinovici, ``Some global aspects of
string compactifications,'' Hebrew University preprint RI-143-90.}
\lref\RSS{M. Ro\v cek, K. Schoutens, and A.
Sevrin, ``Off-shell WZW models in extended superspace,'' IAS preprint
IASSNS-HEP-91-14.}
\lref\CHS{C. Callan, J. Harvey and A. Strominger, ``Worldsheet
approach to
heterotic solitons and instantons,''\ajou Nucl. Phys. &B359 (91) 611.}
\lref\HoSt{G. Horowitz and A. Strominger, `` Black strings and
$p$-branes,''\ajou Nucl. Phys. &B360 (91) 197.}
\lref\worm{S. Coleman, ``Black holes as red herrings: Topological
fluctuations and the loss of quantum coherence,''\ajou Nucl. Phys. &B307
(88)
867\semi S.B. Giddings and A. Strominger, ``Loss of incoherence and
determination of coupling constants in quantum gravity,''\ajou Nucl. Phys.
&B307 (88) 854.}
\lref\BaNe{I. Bars and D. Nemeschansky, ``String propagation in backgrounds
with curved space-time,''\ajou Nucl. Phys. &B348 (91) 89.}
\lref\Bars{I. Bars, ``Heterotic superstring vacua in four dimensions based
on non-compact affine current algebras,''\ajou Nucl. Phys. &B334 (90) 125.}
\lref\MSW{G. Mandal, A Sengupta, and S. Wadia, ``Classical solutions of 2d
string theory,'' IAS preprint IASSNS-HEP-91/10.}
\lref\MaSh{E.J. Martinec and S.L. Shatashvili, ``Black hole physics and
Liouville theory,'' Chicago preprint EFI-91-22.}
\lref\DVV{R. Dijkgraaf, H. Verlinde, and E. Verlinde, ``String propagation
in a
black hole geometry,'' Princeton/IAS preprint PUPT-1252=IASSNS-HEP-91/22.}
\lref\Hawk{S. W. Hawking, ``Particle creation by black holes,''
\ajou Comm. Math. Phys. &43 (75) 199.}
\lref\hair{M. Bowick, S.B. Giddings, J. Harvey, G. Horowitz and A.
Strominger,
``Axionic black holes and an Aharonov-Bohm effect for strings,''
\ajou Phys. Rev. Lett. &61 (88) 2823\semi L. Krauss and F. Wilczek,
``Discrete gauge symmetry in continuum theories,''
\ajou Phys. Rev. Lett. &62 (89) 1221.}
\Title{\vbox{\baselineskip12pt\hbox{UCSB-TH-92-01}\hbox{hepth@xxx/9202004}}}
{\vbox{\centerline{Dynamics of}
\vskip2pt\centerline{Extremal Black Holes}}}

\centerline{{\ticp Steven B. Giddings\footnote{$^\dagger$}{Internet:
giddings@denali.physics.ucsb.edu} and Andrew
Strominger\footnote{$^\ddagger$}{Bitnet: andy@voodoo}}}
\vskip.1in
\centerline{ Department of Physics}
\centerline{ University of California}
\centerline{ Santa Barbara, CA 93106}
\bigskip
\centerline{\bf Abstract}
Particle scattering and radiation by a magnetically charged, dilatonic black
hole is
investigated near the extremal
limit at which the mass is a constant times the charge.
Near this limit a neighborhood of the horizon of the
black hole is closely approximated by a trivial
product of a two-dimensional black hole with a sphere.
This is shown to imply that the scattering of long-wavelength
particles can be
described by a (previously analyzed)
two-dimensional effective field theory, and is
related to the formation/evaporation of two-dimensional black holes.
The scattering proceeds via particle capture followed by Hawking
re-emission, and na\"\i vely appears to violate unitarity.
However this conclusion can be altered when the effects of backreaction
are included. Particle-hole scattering is discussed in the light of a recent
analysis
of the two-dimensional backreaction problem. It is argued that the
quantum mechanical possibility
of scattering off of extremal black holes implies the potential existence
of additional quantum numbers - referred
to as ``quantum whiskers'' -
characterizing the black hole.
%
\Date{}
\newsec{Introduction}

Consider the theory consisting
of Einstein gravity coupled to electromagnetism.
This theory contains charged black hole solutions for which the mass
$M$ equals or exceeds, in Planck units, the charge $Q$. For $M < Q$, a
naked singularity appears.  According to Hawking\refs{\Hawk}, a
quantum mechanical black hole of mass
$M>Q$ will evaporate incoherently until it reaches the extremal value
$M=Q$, at which point the Hawking temperature vanishes and the
evaporation ceases. Thus the extremal solutions are expected to be
the endpoints of Hawking evaporation and correspond to
stable quantum groundstates.

Let us now consider throwing a long-wavelength
particle into the extremal black hole. This results in a non-extremal
black hole with a non-zero Hawking temperature. It should therefore decay
back to (one of) its extremal groundstate(s) via
Hawking emission.\foot{Although for $Q>1$ it might also decay to several
extremal black holes. We shall ignore this possibility in the following.} This
raises the following
interesting question: how does one describe the scattering of
long-wavelength particles in the
presence of an extremal black hole?
A  na\"\i ve application of Hawking's calculation suggests that it cannot
be described by a
unitary $S$-matrix, but rather should follow from a non-factorizable (but
probability-conserving) ``$\Ssl$-matrix''\dollar\ mapping density
matrices to
density matrices.  On the other hand,
the stability of such objects suggests that their scattering might be
similar to that of an elementary particle; indeed many
have speculated
that in a strong sense extremal black holes are equivalent to elementary
particles.

As emphasized
by Preskill {\it et.~al.}\refs{\PSSTW} the semiclassical
methods used by Hawking to estimate decay rates of highly
non-extremal black holes break down when applied to this problem. The
reason for this is that the backreaction of the emitted radiation
on the black hole
inevitably becomes very large near the extremal limit. Consequently
new approximation methods must be found to describe particle-hole
scattering.

In this paper we investigate some basic features of particle capture by
charged dilatonic black holes (such as are found in string theory)
followed by re-emission. It is also true in this context that
near
extremality a discussion of the Hawking process is incomplete without
including the effects of the backreaction of the emitted particles on the
black hole.\foot{This is, however, for slightly
different reasons than those
given in \PSSTW. Unlike the Reissner-Nordstrom case, the temperature
of a charged dilatonic black hole does not
rapidly vary near extremality. The backreaction is nevertheless
important because energy conservation implies a large change in the
geometry during a typical emission process.}  Several features of the
dilatonic black holes render them more amenable to analysis.
Firstly, as stressed
in \GHS, the extremal, magnetically charged, dilatonic black
holes are (unlike their
Reissner-Nordstrom cousins) completely non-singular.
The upper bound on the curvature can be made arbitrarily small by
increasing
the charge. Thus there is no reason to believe that short distance
physics plays any role in low-energy particle-hole scattering.
This strongly suggests that {\it the scattering of low-energy particles
off of extremal black holes is
essentially a problem in low-energy quantum gravity, and is
independent of the cutoff.}\foot{This is less evident, but
we suspect nevertheless true, in the Reissner-Nordstrom case because the
extremal solutions are singular. We also note that this conclusion
could be affected if there is a tendency for large $Q$ black holes to
quantum mechanically fracture to minimal $Q$ black holes, for which the
curvature is large and quantum gravity effects may be important.} Therefore
the role played by string theory in the
present developments is {\it not} to provide a consistent cutoff for
quantum gravity, but rather to suggest modifications of the low-energy
theory
which render the computations more tractable.

Secondly, we shall see that
the low-energy dynamics of near-extremal dilatonic black holes
reduces in a simple way to a two-dimensional problem.
This allows us to introduce a two-dimensional effective
theory to summarize the essential physics.
Progress in understanding this two-dimensional problem, and in particular
in treating the backreaction, was
recently made in \evan, and will be applied to higher-dimensional physics in
the present work.

Our understanding of the two-dimensional problem is
unfortunately insufficient to answer the basic question of
whether or not there is loss
of coherence in scattering from black holes. The discussion of \evan,
translated into the present higher-dimensional context, suggested the
possibility
that the extremal black hole
behaves like an ordinary quantum system with a possibly infinite
groundstate degeneracy.\foot{However very recent work \refs{\RST, \BDDL}
rules out some of the conjectures made in \evan, at least in the form given
therein. The relevance of
\refs{\RST, \BDDL} for the present work will be discussed in section 6.}
The existence of a unitary $S$-matrix then follows if the
number of groundstates is finite.
However if the groundstate degeneracy is infinite, scattering $may$
still lead in practice to quantum incoherence because the quantum state of the
black hole is not observable. However, it is
argued that even in this case there can be
superselection sectors leading to coherent scattering. This must be
so if the
accessible black hole entropy is finite, as indicated by Hawking's area law.
The superselection sectors are labeled by conserved quantum numbers which
are examples of quantities we
refer to as `quantum whiskers.' These are a potentially infinite
set of new parameters  which are
determined by direct scattering off of the black hole groundstate.

This paper is organized as follows. Section 2 contains a description of the
near-extremal
geometry of four- and five-dimensional dilatonic black holes, reviewing
and extending results
of \refs{\Gibb,\CHS,\GHS,\HoSt,\GiSt}. In Section 3 we explain how
coupling to the dilaton imparts an effective mass to most modes in
a large region surrounding the black hole. Section 4 contains
a discussion of semiclassical Hawking radiation in a regime
for which the backreaction can be ignored, and a brief comment on
stringy black holes at or above the Hagedorn temperature. In section 5
a two-dimensional effective field theory is derived for the the description
of low-lying excitations of the extremal black holes. Section 6 addresses
the issue of quantum coherence of particle-hole scattering. The
implications of the two-dimensional analysis of \evan\ are
discussed, and the notion of quantum whiskers is explained. Brief
concluding comments are made in Section 7.

Although there are occasional references to issues in string
theory throughout the paper, we believe that the
implications of our work extend beyond string theory and
hope to have written this paper in
a manner accessible to non-string theorists.

\newsec{Approaching the Extremal Limit}

We begin by describing the peculiar behavior of dilatonic
black holes near the extremal limit where the mass $M$ equals a
constant times
the charge ${Q}$. As will be seen, the geometry greatly simplifies in this
limit.
We first consider the
case of large ${Q}$ (in Planck units)
and restrict attention to the region near or outside
the horizon. In that case the curvature is everywhere weak and sigma model
perturbation theory should be valid.

The four-dimensional black hole solutions of string theory
described in
\refs{\Gibb,\GHS} have higher-dimensional generalizations as found in \HoSt.
In
particular, the five-dimensional member of this
family of solutions (studied in \refs{\CHS,\DuLu,\GiSt}) has a
simple and interesting structure which make it a natural laboratory, along
with the four-dimensional solutions, for
studying black hole dynamics. We begin with this five-dimensional case.

\subsec{The five-dimensional case}

The five-dimensional black hole can be derived as a solution
of the low-energy effective action for string compactification down to
five dimensions:
\eqn\fivact{S_5 =
\int d^5 x \sqrt{-g}
 e^{-2 \phi} \left(
                R + 4(\nabla \phi)^2 - {1 \over 3} H^2
 \right), }
where $H=dB$ is an antisymmetric tensor field strength, fields
which do not enter into the solution are omitted from \fivact,
and we have set $\alpha^\prime = 1$.
The solution can be expressed as \refs{\HoSt}\foot{The following metric is
a simple coordinate transformation of the expression found in \HoSt.}
\eqn\newzmet{ \eqalign{ds^2& = -Q\tanh^2 \sigma d\tau^2
+ \left[\sqrt{Q^2 + {\Delta_5^2}}+{\Delta_5}\cosh 2\sigma
\right]( d\sigma^2 +
d\Omega_3^2)\ ,\cr
e^{2\left(\phi- {\phi}_0\right)} &=  {\sqrt{Q^2 + {\Delta_5^2}}
+{\Delta_5}\cosh 2\sigma
 \over2 \Delta_5 \cosh^{2} \sigma }\ ,\cr
H&=Q \epsilon_3\ ,}}
where the quantity $\Delta_5$ is related to the
mass\foot{Note that the present definition of the mass differs from \GiSt\
but agrees with \HoSt.} $M$
(in the $\sigma$-model metric $g$) by
\eqn\dlta{
\Delta_5 = M -\sqrt{{M^2\over4} + {Q^2\over3}}\ ;}
$d\Omega_3^2$ and $\epsilon_3$ are the line element
and volume form on the unit three sphere;
and $Q$ is integrally quantized in units of $\alpha^\prime$.
A black
fivebrane solution of ten-dimensional string theory may be obtained from
\newzmet\ by simply tensoring with the flat five-dimensional metric.

The extremal limit is
\eqn\limit{
{\Delta_5}{\rightarrow}{0,}
}
\noindent which implies $M=2Q/3$. Near this limit one can distinguish four
regions of the black hole (see Fig.~1):
\relax
\settabs 3\columns
\+&&\cr
\+i)&$\sigma>>{1\over2}\ln ({Q}/{\Delta_5})$&AF region\cr
\+&&\cr
\+ii)&$\sigma\sim{1\over2}\ln ({Q}/{\Delta_5})$&mouth\cr
\+&&\cr
\+iii)&${1\over2}\ln ({Q}/{\Delta_5})>>\sigma>>1$&throat\cr
\+&&\cr
\+iv)&$\sigma=0$&horizon.\cr
\+&&\cr

Region (i) is far from the black hole where the
metric is nearly flat and the dilaton nearly constant.
Region (ii) is the mouth of the black hole at which the curvature becomes
large. At the mouth one enters the long throat region. The proper
length of the throat region is
\eqn\lngth{
{D_T}{\sim}{{\sqrt{Q}}\over{2}} \ln ({Q}/{\Delta_5})
}
\noindent which diverges as ${\Delta_5}{\rightarrow}{0}$. The dilaton
varies nearly linearly along the throat, and the radius of the
three-spheres of constant $\sigma$ and $\tau$ is nearly constant. The throat
then ends at the horizon. The coordinate system~\newzmet~ does not cover
the region inside the horizon, where in any case sigma model perturbation
theory breaks down.

The distinguishing feature of this geometry, shared by the four-dimensional
black hole discussed in the next subsection,
is that the distance from the mouth
to the horizon diverges in the extremal limit. This will be seen to have
interesting physical consequences. It also leads to several
inequivalent ways of approaching the extremal limit. In the first approach,
one keeps the AF and mouth regions (i) and (ii) fixed, while allowing the
horizon to move off to infinity as ${\Delta_5}{\rightarrow}{0}$. The
geometry is then \CHS

\eqn\met{\eqalign{ds^2& =
-Qd\tau^2 + \left(1+{Q \over y^2}
\right)\left[dy^2
+y^2 d \Omega_3^2\right]  , \cr
e^{2(\phi-\phi_0)} &= 1+{Q\over y^2}\ ,\cr
H&=Q \epsilon_3\ ,}}
where $y=\sqrt{2\Delta_5} {\rm cosh}\sigma$.
This describes a supersymmetric ``black hole'' with no horizon or
singularity. Rather there is a semi-infinite throat attached to the AF
region. The dilaton grows linearly and therefore the action \fivact\
becomes strongly coupled far down the
throat. As argued in \refs{\CHS}, in string theory there is an enhanced (4,4)
worldsheet supersymmetry
in this limit, which implies a non-renormalization theorem. Thus ~\met~
describes an exact solution to the classical string equations of
motion\refs{\CHS\DuLu}.

A second method of approaching the limit\refs{\GiSt}
is to keep the horizon and $e^{-2\hat\phi_0}=2\Delta_5 e^{-2\phi_0}$ fixed
while letting the AF region move to infinity as ${\Delta_5}{\rightarrow}
{0}$. One then obtains
\eqn\newmet{\eqalign{ds^2& = -Q\tanh^2 \sigma d\tau^2 + Q d\sigma^2 + Q
d\Omega_3^2 \ ,\cr
e^{2\left(\phi- {\hat \phi}_0\right)} &= Q \cosh^{-2} \sigma\ ,\cr
H&=Q \epsilon_3\ .}}
which describes the horizon region attached to a
semi-infinite throat. The dilaton again grows linearly along the
throat, but approaches weak coupling at the end.
This limit can also be shown\GiSt\ to correspond to an exact classical
string solution by constructing the underlying
conformal field theory as the tensor product of SU(1,1)/U(1) and SU(2) WZW
theories.
This construction
corresponds to a spacetime
which includes the region inside the horizon and near the ``singularity''.

It is also of interest to consider the ``throat limit'' in which both
the horizon and the AF region tend to infinity, leaving an infinite throat
described by
\eqn\wormsol{\eqalign{ds^2&= -Qd\tau^2 +Qd\sigma^2 + Q d\Omega_3^2,
\cr \phi &
=-\sigma + \phit
\cr H &=Q\epsilon_3,} }
with $\phit = \hat\phi_0 + {\rm ln}(2\sqrt{Q})$.
At one end the linear dilaton is strongly coupled, while at the other
it is weakly coupled. This limit also corresponds to a string
solution\refs{\CHS}, given by a level
${Q}$, SU(2) WZW model together with a Feigin-Fuks-like theory.


\subsec{The four-dimensional case}

The extremal limit of the four-dimensional magnetically
charged black hole of
\refs{\Gibb,\GHS} has behavior similar to that of the five-dimensional
black hole of the previous section.
It is a solution of the effective action
arising in string compactification to four dimensions,
\eqn\fouact{S_4 =
\int d^4 x \sqrt{-g}
e^{-2 \phi} \left(
                R + 4(\nabla \phi)^2 - {1\over2} F^2
 \right), }
where $F$ is the electromagnetic field strength and irrelevant terms are
again omitted.  The black hole is given by
\eqn\fdsoln{\eqalign{
ds^2&= -4Q^2 \tanh^2 \sigma d\tau^2
 +\left[2M + \Delta_4\sinh^2\sigma\right]^2
 \left[4d\sigma^2 + d\Omega_2^2\right]\cr
e^{2(\phi-\phi_0)}&= {2M + \Delta_4 \sinh^2\sigma
 \over \Delta_4 \cosh^2
\sigma} \cr F&={Q} \epsilon_2\ ,
}}
where and
$d\Omega_2^2$ and $\epsilon_2$ are the line element
and volume form on the unit two sphere. $M=e^{-\phi_0}M_{ADM}$
is again the sigma-model mass, the magnetic charge $Q$ is quantized, and
$\Delta_4$ is given by
\eqn\dlt{{\Delta_4} = 2M-{Q^2 \over 2M} .}
As previously, in the extremal limit $M\rightarrow Q/2 $ there are four
regions:
\relax
\settabs 3\columns
\+&&\cr
\+i)&$\sigma\gg \half \ln (Q/\Delta_4)$&AF region\cr
\+&&\cr
\+ii)&$\sigma\sim  \half \ln (Q/\Delta_4)$&mouth\cr
\+&&\cr
\+iii)&$\half \ln (Q/\Delta_4)\gg \sigma \gg 1   $&throat\cr
\+&&\cr
\+iv)&$\sigma= 0$&horizon.\cr
\+&&\cr

In the extremal limit the throat approaches infinite length as before.
Therefore at $\Delta_4=0$ there are three distinct solutions.
The AF region plus infinite throat is given by
\eqn\fdaf{\eqalign{ds^2& = -4Q^2d\tau^2 +\left(1+{Q \over  y}
\right)^2\left[dy^2 +y^2 d \Omega^2\right]\ , \cr
e^{2(\phi-\phi_0)} &= 1+{Q\over  y}\ ,\cr
F&={Q} \epsilon_2\ .}}
Unlike the five-dimensional case, this
extremal limit is not supersymmetric\GHS.
The horizon plus infinite throat is
\eqn\fdhor{\eqalign{ds^2 &= -4Q^2 \tanh^2 \sigma
d\tau^2 +  4Q^2
d\sigma^2 +
 {Q^2} d\Omega_2^2\ ,\cr
e^{2(\phi-\hat\phi_0)}& = { Q\over \cosh^2 \sigma }\ ,\cr
F&={ Q} \epsilon_2\ .}}
Note that this solution is a product of two two-dimensional solutions, and that
again one of them is the low-energy limit of the exact black hole solution
of \Witt.  This means that the extremal four-dimensional black hole should
correspond to an exact solution of string theory which is the product of
two conformal field theories\GiSt.  The theory describing the angular
variables corresponds to the round two sphere penetrated by uniform
magnetic flux.   Finally, there is the limit corresponding to the
infinite throat with linear dilaton as in \wormsol :
\eqn\fthr{\eqalign{ds^2&= -4Q^2d\tau^2 +4Q^2d\sigma^2 + {Q^2}
d\Omega_2^2,
\cr \phi &
=-\sigma + \phit
\cr F &=Q\epsilon_2,} }

We note that at the classical level there are also electrically
charged\refs{\Gibb,\GHS} as
well as dyonic solutions\refs{\Gibb,\ddd}.
However in a theory with massless charged
fermions, such as heterotic string theory, the electric charge will be
rapidly discharged through
Schwinger pair production.

\newsec{Perturbations on the throat}

We have argued that near the extremal limit the essential features of the
geometry are the asymptotically flat region and the attached long throat.
For the purpose of
studying emission and absorption of particles
it is important to elucidate propagation of fields along this
throat.  In particular, in the Hawking process particles are emitted in the
vicinity of the horizon and must propagate up the throat to escape to
infinity.  In this section we will argue that interaction with the linear
dilaton results in significant
attenuation for most low-energy particles.  (This simple conclusion was
also independently derived in the more detailed analysis of \HoWi; there
the argument was made in the `canonical' metric, which differs from the
metric herein by a dilaton rescaling.)

This propagation is described by considering small perturbations of the
fields about the linear-dilaton solution; the equations governing these
perturbations are obtained by linearizing the equations arising from the
actions \fivact, \fouact, or their generalizations incorporating other
fields.  Although straightforward, this
is somewhat complicated due to mixing between the perturbations, \etc\
However, for perturbations moving along the throat,
the couplings to the background
are quite simple:  the metric in the direction along the throat is flat,
and there is an exponential coupling to the linear dilaton.  Thus when one
diagonalizes the kinetic matrix one finds actions of the form
\eqn\SFact{S_\psi = -\int d^{D}x\sqrt{-g} e^{-2b\phi}\left(\nabla
 \psi\right)^2\ }
governing the generic perturbation $\psi$, and the constant $b$ depends on
the
mode in question.  Were it not for the dilaton background this
action would describe free, massless propagation along the throat.
However, the linear dilaton modifies the dispersion relation.  This can be
seen by defining a new
(canonically normalized) fluctuation  by
\eqn\newf{\psih =e^{-b\phi}\psi\ .}
One then finds
\eqn\mtact{S=-\int d^Dx\sqrt{-g}\left( (\nabla \tilde \psi)^2+b^2(\nabla
\phi)^2
\tilde \psi^2
+b\nabla\phi\nabla \tilde \psi^2\right)\ .}
Along the throat the last term is a total derivative and
the effect of the linearly varying dilaton is simply to give a mass
$m^2=b^2\left(\nabla\phi\right)^2$ to
$\tilde \psi$,
\eqn\THact{ S_{\rm throat} = -\int d^{D}x\sqrt{-g} \left[\left(\nabla
\psih\right)^2 +m^2 \psih^2\right]\ .}
The mass is $m = b/\sqrt{Q}$ in the five-dimensional case and $m=b/2Q$ in the
four-dimensional
case.  Propagation of
excitations below the mass gap is exponentially suppressed.
This is in accord with the observation of \refs{\GHS} that there
is a barrier surrounding the near-extremal black
hole that suppresses emission
and absorption.

\newsec{Semiclassical evaporation near the extremal limit}

A dilatonic  black hole formed from
gravitational collapse will in general emit
ordinary Hawking radiation.  Holes with arbitrary mass
$M>M_{\rm extremal}$
will thus eventually approach the extremal limit.  In this section we
will use semiclassical methods to
investigate the properties of the radiation near the extremal limit. As
will be
quantified shortly, these methods break down before the extremal limit is
reached due to a large backreaction.
However, for large $Q$, there is nevertheless a range of near-extremal
values of $M$ for which the backreaction is negligible.
As we will see, in this range the radiation has some rather novel features.

The inverse temperatures
\eqn\Temp{ \beta_4 = 8\pi M\quad ,\quad \beta_5 = 2\pi\left(\sqrt{Q^2 +
\Delta_5^2} + \Delta_5\right)^{1/2}
}
of the black holes
are given by the periodicity of the euclidean sections.  Note
that as the extremal limit is approached
these temperatures approach finite values
\eqn\Ext{T_4 = 1/4\pi Q\quad ,\quad T_5=1/2\pi\sqrt{Q}  \ .}
However {\it at} the extremal limit, no identification is required to make
the Euclidean section regular, and the temperature vanishes.

\subsec{Radiation for $Q\gg1$}

Consider the near-extremal range
\eqn\Nxl{\eqalign{Q&\gg\Delta_4\gg 1/Q\quad (D=4)\cr
                  Q&\gg\Delta_5\gg 1/\sqrt{Q}\quad (D=5)}
}
For $\Delta$ less than the upper bound the length of the
throat exceeds its width. The lower bound is required
in order that the backreaction can be neglected.
The backreaction is
important whenever the energy of an emitted quantum is comparable
to the deviation of the mass from extremality since then the geometry
changes significantly during the emission of the quantum.
Near extremality this deviation is proportional to
$\Delta$, so the backreaction becomes important when
$\Delta\sim T$; the temperature relations \Ext\ then give the
lower limits in \Nxl.
The important case where $\Delta$ is less than this
bound will be discussed in Section 5.

In order that the range \Nxl\ be large we need
$Q\gg1$. This also implies that the temperature is small as compared to the
Planck temperature, $T\ll1$, and that, within the context of string theory,
higher mass string modes may be ignored.

As argued in the preceding section,
perturbations propagating along the dilaton throat into or out of the hole
are governed by the action \THact.
The radiation rate for such perturbations can be calculated by the usual
Hawking\refs{\Hawk} methods.  This rate may be estimated for small $\Delta_4$
as follows. (We describe the four-dimensional case, although the
five-dimensional case is similar.)
The number of particles emitted from the
vicinity of the horizon
of the hole per unit time in the energy range $(E, E+ dE)$ should be
governed by the usual thermal factor
\eqn\thermal{dn= {dE\over 2\pi} {1\over e^{\beta E} \mp 1}}
(with $\mp$ for bosons or fermions) times the transmission coefficient for
a particle of energy $E$ to escape from the vicinity of the
horizon to infinity.  In the case of an ordinary black hole this latter
coefficient is of order unity; however, the effective mass in \THact\
modifies this behavior near extremality.
Indeed, propagation of excitations below the mass gap, $E<b/2 Q$,
is exponentially suppressed with amplitude
\eqn\Amp{A(\sigma)= \exp\left\{- \sqrt{b^2 -4Q^2 E^2} \sigma\right\}\ ,}
Since the throat has extent
$\sim {\rm ln}\left(Q/\Delta\right)$, this means that the tunneling
rate has a suppression factor $\sim \Delta^{b}$ for energies far below the
gap.\foot{This can equivalently be seen without the field redefinition
\newf\ if one takes into account the factor $e^{-2b\phi}$ in the
normalization.}

This is not, however, sufficient to imply the vanishing of the radiation
rate in the extremal limit.  The reason is that the temperature
in this limit is $T\sim1/{Q}$ and therefore has appreciable
probability for producing particles with energies
above the mass gap.  Since the proper
radius of the mouth is also of order $Q$, the energy needed for a
particle of angular momentum $l$
to be transmitted into or out of the throat is
given by $E^2\sim b^2/4Q^2 + l(l+1)/Q^2$.  Therefore the
s-wave ($l=0$) dominates.
A rough estimate of the evaporation rate can then be made by
approximating the transmission coefficient by unity
for $E^2 > \left[b^2/4 +l(l+1)\right]/Q^2$ and zero otherwise.  Then the sum of
transmission probabilities over angular momenta is approximately
\eqn\transmit{T(b,E) \simeq \sum_{l} (2l+1)\Theta\left(\sqrt{4Q^2E^2 - b^2} -
\sqrt{l(l+1)}\right) \sim 4Q^2E^2 -b^2}
for $E>b/2 Q$.  This yields an estimate of the rate loss of mass from
the hole:
\eqn\massloss{{dM \over dt} \sim - \int_{b/2 Q}^\infty {EdE\over
e^{4 \pi Q E} \mp 1} \left(4Q^2E^2 -b^2\right) \sim -{1\over Q^2}\ .}
Note that the rate approaches a nonzero value in the extremal limit.  (This
is contrary to \PSSTW\ but in accord with the independent observations of
\HoWi.)

This na\"\i ve estimate might seem to suggest that
evaporation could continue beyond extremality to produce a naked
singularity.  However,
a crucial
ingredient has been neglected: the backreaction of the radiation on the
geometry of the black hole.  As stated above, this will become relevant when
$\Delta_4\sim 1/Q$ $\left(\Delta_5\sim 1/\sqrt Q\right)$.
It is clearly very important, and difficult, to
investigate the physics of the black hole in this limit to resolve the
issue of what happens in the end stages of the Hawking process.
Although we cannot at present give a full treatment of the
four- or five-dimensional problem of the backreaction, progress can be made
in this direction by exploiting the connection between the
higher-dimensional solutions and the two-dimensional black hole.  This will
be discussed in section 5.

\subsec{Radiation for $Q\sim 1$}


Now we turn briefly to the question of Hawking radiation for
near-extremal five-dimensional black holes with small values of $Q=n$.
(Although similar statements could in principle be made for the
four-dimensional holes, the exact form of these solutions for small $n$ is
not known.)
One might
expect that quantum effects will
allow extremal black holes to bifurcate. For the five-dimensional case
supersymmetry suggests that higher $Q$ black holes are neutrally stable,
but they might be split up into lower $Q$ black holes by particle
scattering.
If this is the case,
$Q=1$ black holes would behave the most like fundamental particles
and would be the most
interesting objects to study. However it is clear that
quantum gravity effects are then important. In the context of string theory
the low-energy field theory approximation is not valid because
the Hawking temperature near $\Delta_5=0$,
\eqn\temp{ T_5
\simeq {1 \over 2\pi\sqrt{n\apm}}\ , }
is
high enough to excite massive string modes. (Here we have momentarily
restored the suppressed factor of $\apm$.)  Indeed the Hagedorn
temperatures are
\eqn\hag{T_c={1 \over 2\pi\sqrt{2 \apm}}~{\rm (type~II)},~~~~~T_c={1
\over(1+\sqrt 2 )\pi\sqrt{2\apm}}~{\rm (heterotic)}.}
Thus in the heterotic case the $n=1,2$ black holes are above the Hagedorn
temperature in the heterotic theory, as is the $n=1$ hole in the type II
theory.  Furthermore, the $n=2$ hole is precisely {\it at} the Hagedorn
temperature in the type II theory.
This obviously brings some new and fascinating issues
- beyond the scope of the present work -
into the study of extremal black holes.  Recent work
\refs{\CHS,\Witt,\GiSt} obtaining the five-dimensional extremal black holes
as {\it exact}
solutions of string theory may be useful in this regard.

\newsec{The Low-Energy Effective Field Theory}

In the preceding section we have argued that very near extremality,
$\Delta<T$, the backreaction must be included in examining the
subsequent evaporation of the black hole. Though
the dynamics in this region are not well-understood, it
is reasonable to suppose that
the energy of an emitted quanta in this region is bounded by $M-M_{\rm
extremal}$,
since the emission of a quanta exceeding this bound would result in a naked
singularity. For sufficiently small $M-M_{\rm extremal}$, this
implies an exponential suppression for
emission of modes which are
massive
along the throat. The dynamics are then dominated by those modes which are
massless along the
throat, i.e. have $b=0$ in \SFact. As we shall see, such massless modes do
arise in string theory,
despite the generic tendency for dilaton-induced masses.

The regime $\Delta\ll T$ is also the regime
of interest for the problem discussed in the introduction of scattering
particles off of an extremal black hole. If the energy of the incoming
particle is sufficiently low, the extremal black hole can never be excited
to a state with $\Delta_4 \sim 1/Q$ ($\Delta_5\sim 1/\sqrt Q$).
The hope is that particle-hole scattering
can be described by perturbation
theory about the extremal black hole.

Since all the relevant dynamics for $\Delta_4\ll 1/Q\,$
($\Delta_5\ll 1/\sqrt Q$)
occurs at length scales much greater than $Q\,$ ($\sqrt Q$), it is useful to
derive a low-energy effective action to describe these dynamics.
The backreaction problem may then be analyzed in this context.
It will turn out that this effective action is
partially two-dimensional, and leads to a direct and simple relation
between the process of absorption and re-emmision of a particle by
a higher-dimensional extremal black hole with the process of
formation and re-evaporation of a two-dimensional black hole.
%
%
%

\subsec{The Throat Limit}

Let us first consider the throat limits \fthr\ (\wormsol ) of the four-
(five-) dimensional
black holes.  In that limit the geometry is the product of two
dimensional Minkowski space with a two-(three-)
sphere of radius $Q$ ($\sqrt Q$). The effective action at
scales longer than $Q$ ($\sqrt Q$) can be derived by the usual Kaluza-Klein
procedure for compactification from four (five) to two dimensions on a two-
(three-) sphere.
The result is
\eqn\twoact{S_2 =
\int d^2\sigma \sqrt{-g}
 e^{-2 \phi} \left(
                R + 4(\nabla \phi)^2 +4\lambda^2
 \right), }
with $\lambda^2 = 1/4Q^2\,$ (1/Q) for four dimensions (five dimensions).
In addition there are two-dimensional gauge fields arising both as
relics of higher-dimensional gauge fields and
from isometries of the
compactification
spheres. These gauge fields may lead to dyons and other
interesting effects
(they do not propagate in two dimensions), but it is
consistent to set them to zero as shall be done in the present analysis.
Corrections to \twoact\ are
suppressed by powers of $Q$.

Eq.~\twoact\ has a two-dimensional black
hole solution given by
\eqn\twohole{\eqalign{ds^2& = -\lambda^{-2}\tanh^2 \sigma d\tau^2 +
\lambda^{-2}
d\sigma^2  \ ,\cr
e^{-2\left(\phi- {\hat \phi}_0\right)} &= \lambda^2 \cosh^{2} \sigma\ .}}
which is precisely the two-dimensional portion of the higher-dimensional
black hole solutions described in the previous sections (as well as,
for $\lambda^2=k/2$, the
low-energy limit of the exact conformal field theory discussed in \Witt).
Thus the effective action \twoact\ actually
describes all the configurations of the
horizon limit.

One key to the relation between the two- and higher-dimensional scattering
problems alluded to above is now apparent.
The masses of the four-dimensional and five-dimensional black holes are
related to the value of the dilaton at the
horizon by
\eqn\dilh{e^{-2\phi_h} = {2 e^{-2\phi_0}\Delta_4\over\sqrt{4Q^2 +
\Delta_4^2} + \Delta_4}\quad {\rm and}\quad
e^{-2\phi_h} = {2 e^{-2\phi_0}\Delta_5\over\sqrt{Q^2 +
\Delta_5^2} + \Delta_5} .}
from \newzmet\ and \fdsoln.
But recall\Witt\ that the mass of the two-dimensional
black hole is given by the value of the
dilaton at the horizon:
\eqn\tdmass{M_{2d} \propto e^{-2\phi_h} \propto  e^{-2\phi_0}{\Delta_{4,5}\over
Q}\ }
where we work near the extremal limit.
Thus the evaporation of or scattering by higher-dimensional black holes
near extremality is closely related to analogous processes for
two-dimensional black
holes near zero
mass.\foot{This is in accord with the suggestion of
E.~Witten\refs{\Witt,\WittSton} that the two-dimensional vacuum
be interpreted as an
extremal black hole.}

While \twoact\ does have
interesting  black hole solutions, it is a theory with no
propagating degrees of freedom; the two constraints and two gauge
conditions
of two-dimensional gravity eat up the $3+1$ degrees of freedom of the metric
plus
dilaton.
This means that gravitational collapse and Hawking radiation can not be
studied in the theory \twoact. However, within the context of an enlarged
theory such as string theory, there are
additional massless fields, neglected in \fouact\
and \fivact, which do lead to
propagating degrees of freedom in \twoact. The precise form of these
fields differs between the four and five-dimensional case, and also depends
on the type of
string theory under consideration.

For
a four-dimensional compactification of heterotic
string theory there are for example typically massless scalars
$M$ resulting from compactification moduli. These are
governed by an action of the form
\eqn\modact{S_M=\int d^4x\sqrt{-g}e^{-2\phi}(\nabla M)^2,}
(compare \SFact).
This  should be added to \fouact.
The factor of $e^{-2\phi}$ appears in front
of all terms in the classical action. As in \newf-\THact\
it may be eliminated by the field
redefinition $\hat M=e^{-\phi}M$.
Would-be massless
contributions to the two-dimensional effective action arise from modes of $M$
which are constant on the compactification two-sphere.
However, as in \THact\
it is clear
that even this lowest-energy mode does {\it not} give rise to a
massless two-dimensional mode, since the dilaton background produces a mass
of order ${1/Q}$ ($1/\sqrt Q$).  More generally, massless
bosons in four dimensions with the $e^{-2\phi}$ prefactor
will not lead to massless two-dimensional fields.

The situation is different in some cases for fermions.
The effective action for a four-dimensional heterotic string
compactification contains the charged fermion term
\eqn\fhet{S_\chi=\int d^4x\sqrt{-g}e^{-2\phi}{\bar \chi}
\overleftrightarrow\Dsl
\chi,}
where $D$ here is the gauge covariant derivative. The $e^{-2\phi}$
prefactor may be absorbed by defining ${\hat\chi}= e^{-\phi}\chi$;
\eqn\fht{S_\chi=\int d^4x\sqrt{-g}{\bar{\hat\chi}} \overleftrightarrow\Dsl
{\hat\chi},}
Unlike the case of bosons, this redefinition does not produce an effective
mass term.\foot{We are
grateful to Mark Alford for useful conversations on this issue.}
$S_\chi$ will lead to a massless field in \twoact\ if $\hat\chi$
has a zero mode on
the two-sphere. An index theorem relates the number of such zero modes
to the integral of $F$ over the two sphere.
One finds that \twoact\ can be supplemented by
a charged Dirac fermions $\psi$ containing both chiralities:
\eqn\psiact{S_\psi=\int d^2\sigma\bar \psi \Dsl \psi.}
These can be seen to be the only low-energy modes on the
throat for the four-dimensional black holes.

The situation again differs for the five-dimensional black holes. In this
case the two form field strength $F$ is
replaced by the three-form $H$. However since fermions do not couple
directly to the corresponding potential $B$, there are no
corresponding zero modes. In fact for heterotic string theory it can be
seen that there are no zero modes at all.  This is
not the case for type II strings. For example the type IIa string
contains a Ramond-Ramond four-form field strength ${F} = {dA}$ governed by the
low-energy action
\eqn\ften{
{S_F} = {\int}{d^{10}}x\sqrt{-g}{F_{MNPQ}}{F^{MNPQ}}.}
The usual factor of ${e^{-2\phi}}$
could be reinstated in~\ften~ by redefining ${A}$, but
this would result in an $Ad\phi$ term in ${F}$ and a nonstandard gauge
transformation law for ${A}$. After reducing to five dimensions,
$F$ is equivalent to a single scalar field $f$ defined by
\eqn\fdual{F=*df,}
where $*$ is the Hodge dual. Further reducing to two dimensions on the
three-sphere, one
obtains the effective action:
\eqn\fact{S_f=\int d^2\sigma \sqrt{-g}(\nabla f)^2}
Thus ${f}$ is a massless scalar which moves along the throat.


\subsec{Interacting Effective Field Theory}
In the previous subsection the effective field theory governing
long-wavelength excitations
of the throat was derived. The effective field theory governing the
AF region outside the mouth is given by \fivact\ or \fouact, supplemented
by terms describing the other fields. The
full effective field theory is obtained by gluing the two field theories
together at the black hole mouth, as we now describe.

For notational
simplicity we
consider only the five-dimensional black hole and
only the dynamics of the massless scalar field $f$, treating the metric and
dilaton as fixed backgrounds.
Let us separate the scalar field $f$ into
$f_2$ and $f_5$ which have
support on the throat and asymptotic region respectively.
The scalar field on the throat is governed by the action
\eqn\ftwo{
S_2[f_2] =  {\int}{d}^2{\sigma}({\nabla}{f_2})^2
}
\noindent where ${-}{\infty}<{\tau}<{\infty}$ and ${0}<\sigma<{\infty}$.
The mouth of the throat is at ${\sigma} = {0}$, and boundary conditions
there will be discussed shortly. Outside the black hole, the scalar modes
${f_5}$ are governed by the five-dimensional effective action
\eqn\ffive{
{S_5[f_5]} = {\int}{d^5}{x}({\nabla}{f_5})^2.
}
\noindent Thus the full long-distance action has two pieces: one corresponding
to the region where the universe is effectively five-dimensional, the
other where it is effectively two-dimensional!

The field theories corresponding to the two different regions are matched
along the worldtube ${X^\mu}({\tau})+Rn^\mu(\Omega)$ of the black hole mouth.
(Here $n^\mu(\Omega)$ is the unit normal to ${\rm S}^3$.)
The values of the scalar fields ${f}$
along this worldtube
must agree as the mouth is approached from inside or outside, and this in
general leads to the boundary condition
\eqn\fbc{
f_2(\tau,\sigma=0,\Omega) = f_5(X^\mu(\tau) + Rn^\mu(\Omega))\ .
}
Within the functional integral this functional constraint may be enforced
by Lagrange multipliers.  In the low-energy limit where modes with non-zero
angular momentum are
neglected the mouth is replaced by a point and
\fbc\ reduces to the $\Omega$ independent constraint
\eqn\fbca{f_2(0,\tau) = f_5(X^\mu(\tau))\ .}
This boundary condition is enforced by introducing a
single (time dependent) Lagrange multiplier,
${\beta}({\tau})$, into the functional integral:
\eqn\fint{
{Z} = {\int}{\cal D}{f_2}{\cal D} f_5{\cal D}{\beta}
{e^{iS_5 + iS_2 + iS_\beta}}
}
where ${S_5}$ and ${S_2}$ are given by~\ffive\ and~\ftwo,
and ${S_\beta}$ is given by
\eqn\salpha{
{S_\beta} = {\int}{d}{\tau}{\beta}({\tau}){f_5}
({X}({\tau})) - {\int}{d\tau}{\beta}({\tau}){f_2}
({0,\tau})\ .
}
$Z$ involves a weighted integral over all possible Dirichlet-type
boundary conditions at the boundary $\sigma=0$ of the two-dimensional
field theory.
One can then integrate out both the boundary value $f_2(0,\tau)$ and
$\beta(\tau)$ to reexpress the functional integral
in terms of a two-dimensional field theory with a fixed Dirichlet boundary
condition $f_2(0,\tau)=0$ and operator insertions at $\sigma=0$. The
result is:
\eqn\zint{
{Z} = {\int}{\cal D}_D{f_2}{\cal D} f_5
{e^{iS_5 + iS_2 + iS_I}}
}
where
\eqn\sint{S_I=\int d\tau f_5(X^\mu(\tau))\partial_\sigma f_2(0,\tau)}
and the subscript `$D$' on the integration measure denotes that
$f_2(0,\tau)=0$.

More generally one must include effects due to the finite mouth size.
The technology for including these effects has been developed in the
context of wormhole theory. The most general expression consistent
with locality  and energy conservation is
\eqn\geneff{S_\beta= \int d \tau C_{ia}
 {\cal O}_5^i (X(\tau )) {\cal O}_2^a(0,\tau)}
where the ${\cal O}$'s are  complete sets of local operators in the
theories
and a detailed calculation is required to determine the constants
$C_{ia}$.  The coefficients $C_{ia}$ for the angle-dependent modes will be
suppressed due to the large effective barriers.

This analysis extends in an obvious way to incorporate the other fields in
the theory. $S_2$ and $S_5$ and the measure in \zint\ are promoted to the full
expressions
involving the metric and dilaton. Since (\eg) the dilaton
has no propagating massless excitations along the throat, a low-energy dilaton
pulse
incident on the black hole cannot enter the throat unless it turns into
an $f$-pulse--a process which occurs only at higher orders. The boundary
conditions lead to the important
constraint that the (fixed)
value $\phi_0$ of the dilaton zero mode in the asymptotic region match the
value of the dilaton at the mouth $\sigma=0$ of the two-dimensional region.

While the four-dimensional case is largely similar, some new features arise
due to the well-known peculiarities of the charged Dirac equation in the
presence of a magnetic source. This will be the subject of a separate
publication\alf.

In summary, the low-energy effective action contains an
asymptotically flat four- or five-dimensional and
a two-dimensional piece joined along the black hole mouth. Interactions
between the two
pieces are represented by local operators integrated along the
mouth worldline.

\newsec{Is there a unitary $S$-Matrix?}

We now return to the issue of how to describe scattering from extremal
black holes.  In the present context we may wish to consider, for example,
scattering of low-energy $f$
particles off of extremal black holes.

There are two distinct possibilities that have previously been discussed in
the literature:  either black holes exhibit intrinsically non-unitary
dynamics or they undergo coherent quantum-mechanical evolution.
In the former
case it has been argued\dollar\ that black hole dynamics may be described by a
probability-conserving but non-unitary $\Ssl$-matrix.  Such a matrix
linearly maps density matrices to density matrices, but allows arbitrarily
large loss of information or increase in entropy.

Although non-unitary dynamics is a logical possibility that cannot be
ruled out at present,
there are several reasons to favor the alternative.  One of
these is the classical thermodynamical result that a fixed entropy should
be associated to a black hole of a given mass:  this is a result that would
be expected if at the fundamental level the black hole were a quantum
mechanical system.  Indeed, with the definition $S=-{\rm
Tr}\left(\rho\ln\rho\right)$ of the entropy, an $N$-state quantum system
has a maximum entropy $\ln N$.  This latter result is suggestive that not
only are black holes quantum systems, but that they have finitely many
accessible
states at a given mass level.  (We will examine a different possibility
shortly.)

In the present context this problem naturally divides into two parts.
The first is unitarity of the two-dimensional effective field
theory of \S5.1 (argued there to describe
particle-hole scattering
after the $f$-particle enters the throat region).
If this two-dimensional theory is not unitary (e.g.
due to singularity formation), then
it is highly unlikely that low-energy
$f$-particle-hole scattering is unitary.
On the other hand
unitarity of
the two-dimensional field theory \twoact\ does not
necessarily imply that there is a unitary $S$-matrix for
$f$-particle-hole scattering.
The reason
for this is simple: from the higher-dimensional viewpoint, the state of
the two-dimensional field theory inside the black hole mouth is
unobservable and should be traced over.\foot{This is somewhat different than
the
Reissner-Nordstrom case, where the traced-over
state of the black hole is inaccessible
due to an event horizon, and the consequent
loss of unitarity is perhaps physically less disturbing.}
More explicitly, one first computes the $S$-matrix
for the full $S_5+S_2$ field theory.  This maps the initial density
matrix $\rho(0)=\rho_2(0)\otimes\rho_5(0)$ to the final density matrix
$S\rho(0)S^\dagger$.  One then traces over the unobservable $f_2$
field theory to obtain the five-dimensional density matrix, $\rho_5 =
tr_{2}\left[S\rho(0) S^\dagger\right]$.  This will in
general correspond to a mixed
state.

Some progress on the question of unitarity of the
two-dimensional theory was made
in \evan, in which
the process of black hole formation and evaporation
in the two-dimensional field theory \twoact\ (the two-dimensional
analog of particle-hole scattering) was analyzed.
It was suggested that a collapsing $f$-wave (that is an $f$ wave
heading toward $\sigma=\infty$) dissipates its
energy via Hawking evaporation
and that the classical singularity at $\phi=\infty$ (strong coupling)
is removed.
The absence of any singularities would imply the existence of a unitary
$S$-matrix. However more recent work \refs{\RST, \BDDL} has
provided strong evidence that, while this strong-coupling
singularity may indeed not appear, other types of
singularities occur in the quantum theory at a fixed, critical value
$\phi_c$ of $\phi$. In the linear dilaton vacuum  (which corresponds to the
extremal black hole)
$\phi=\phi_c$
is a timelike line which separates the vacuum into two
regions governed by different dynamics.  The singularities found in
\refs{\RST, \BDDL} occur when
an incoming $f$ particle hits this critical line.\foot{These singularities
were noted in perturbation theory in \evan, but it was not
known if they persisted in the non-linear theory.} The proper interpretation
of these singularities is not evident to us at present. However in
the present context we note that they might be avoided as
follows.\foot{The following idea was independently discussed in \BDDL.}
The description of $f$-particle-hole scattering does not require the
full two-dimensional field theory, since the latter is glued on to
the higher-dimensional theory near the line $\phi=\phi_0$, where
$\phi_0$ is approximately the higher-dimensional asymptotic value of
$\phi$. If we choose $\phi_0 > \phi_c$, then the singular line of the
two-dimensional field theory is avoided entirely. By introducing a large
number $N$ of $f$ particles, this may be accomplished within the
weak coupling regime, and the corresponding scattering processes perhaps
computed
perturbatively.
However it is not clear to us at present that
the problems do not pop up somewhere else, and this suggestion should be
regarded as tentative.


While the resolution of these issues is extremely important, to
proceed for the
moment we assume that the two dimensional field theory is
unitary, and consider the consequences for higher-dimensional
$f$-particle-hole scattering.
In general, whether or not this scattering is effectively unitary depends
both on
the internal theory of the black hole (as given by the
two-dimensional theory of \S5.1) and on the
couplings between this theory and the external world.
There are important constraints on these couplings in the present
context, as follows. The results of \evan\ suggest that the interaction
time scale for $f$-particle-hole scattering
is of the order of the black hole size, i.e. the interaction lagrangian
of the low-energy field theory is local in both time and space.
Therefore the effect of the incident
$f$ particle in the low-energy approximation
is simply to induce transitions among the extremal
black hole ground states. This corresponds to an interaction lagrangian of
the form
\eqn\sct{S_I=\int d \tau C_{ia}{\cal O}^i_5(X(\tau))T^a,}
where the $T^a$ are time-independent operators which act on the
black hole groundstates.\foot{Note that a mixed functional integral-operator
formalism is being used here. Because the operators $T^a$ do not in general
commute with one another and are multiplied by time dependent functions
${\cal O}$, it is necessary to time-order the
evolution generating exponentials.} If the number of groundstates is
$N$ (where $N$ may be infinite) then the $T^a$ can be taken to be the
generators of $U(N)$.

If the number of ground states $N$ is finite, this has the important
consequence that there
can be no real loss of quantum coherence.
In that case the final density matrix, though not that of a pure state,
is of the form which arises in the example
of scattering off of a molecule with $N$ degenerate groundstates whose
quantum state is initially
unknown. This is of course the usual case for real laboratory experiments!
Initially,
scattering experiments which induce transitions among the molecular
groundstates will lead to an increase in the entropy of the
experimental apparatus. However, its
entropy cannot increase to more than $\ln N$.
After many scattering experiments, the quantum state of the molecule
is eventually determined, and further experiments do not then lead to
a further increase in the entropy of the experimental apparatus.
Thus, if the black hole has a finite groundstate degeneracy,
the incoherence which arises from scattering experiments is
no more alarming than that encountered in real laboratory
experiments.

While not much is known about the groundstates, the
possibility of an infinite degeneracy certainly can not be ruled out.
In fact the recent analysis of \evan\ suggests that the black hole
could have an
infinite number of groundstates labeled by different values of a global
conserved charge. While an infinite degeneracy
is a necessary condition for  effective
quantum incoherence (here we mean over and above what ordinarily occurs in the
laboratory),
it is not a sufficient condition, as can be seen from
the
following example. Suppose the infinite groundstates $\ket{m}, m~\in ~Z$
of the black hole are characterized
by $q\ket{m}=m\ket{m}$ where $q$ is an operator,
and that the only coupling to the observable
sector is given by
\eqn\cplg{H_i= f^2(X(\tau))q,}
where $X(\tau)$ is the worldline of the black hole and $f$ is a
quantum field operator which creates $f$ particles.
This theory has a superselection rule forbidding transitions
among the groundstates. Off-diagonal elements of the density matrix can
not be measured, and may be set to zero. The density matrix
is then of the form:
\eqn\hstate{\rho=\sum_m \rho_m\ket{m}\bra{m}.}
This means that the black hole is in the state $\ket{m}$ with probability
$\rho_m$.
The eigenvalue $m$ of $q$, which determines the quantum state of the
black hole, may then be measured (to arbitrary finite accuracy)
in a finite number of $f$-particle
scattering experiments. Once the state is determined, the results of
further scattering
experiments can be predicted with quantum mechanical determinacy.

The eigenvalue of $q$ in this example is a new type of
quantum number which can label the
internal state of an extremal black hole. Classically black holes are
characterized by only a few parameters which are generally locally
conserved charges. Quantum mechanically they
may carry an additional few varieties of
``quantum hair'' which can be measured by long range interference
experiments involving strings\hair. We are finding here that there
exists the possibility of additional
observable parameters characterizing the internal state of the black hole.
These are detectable precisely because
black holes are not black: the parameters determine how an incoming
particle scatters to an outgoing
one. The number of such parameters is potentially infinite when one
considers all possible scattering experiments. Whether there are
sufficiently many of them to completely characterize the internal quantum
state of the black hole depends on the details of the scattering process.
Like quantum hair, these new parameters exist only at the
quantum level. Unlike quantum hair, they can be detected only in a
short range scattering experiment. We therefore refer to them as
``quantum whiskers.''

We wish to emphasize that it is a logical possibility that the quantum
whiskers as specified are determined purely by the hair, either classical
or quantum, on the black hole.  In that case they would not further
distinguish
different black holes. But it is also possible that the
quantum whiskers are independent quantities that
can take on infinitely many values, perhaps providing a
memory of the initial collapsing state which formed the black hole.
A more detailed dynamical analysis is required to answer this question.

%
The coupling given in \cplg\ is of course very special.
The general situation is as follows.
The linear combination
of operators $q_A\equiv H_{iAa}q^a$ appearing in the coupling
$H_i$ of the black hole to the external world form an algebra of
observables that generates a subgroup
$G$ of the group $U(N)$ of unitary
transformations of the $N$-dimensional
(where $N$ may be infinite)
space of black hole groundstates. $G$
need not be simple or finite-dimensional. The space of groundstates
then decomposes into irreducible representations $V_r$ of dimension $d_r$
of $G$. The general density matrix describing the black hole
assigns a probability $\rho_r$ for the black hole to be in the
representation $V_r$ (off-diagonal elements which mix representations can
again be set to zero.) If a finite number of representations occur, the
irreducible representation can be determined (\eg~by measuring the
Casimirs)
after a finite number of experiments.
The irreducible representation is
a conserved quantum whisker labeling the black hole.
(The quantum whiskers corresponding to values of the other observables are
not necessarily conserved, as they may be changed in scattering experiments.)
If the irreducible representation so determined finite-dimensional,
the subsequent increase in entropy of the experimental apparatus is bounded
as before by $\ln~ d_r$. The quantum state of the system can eventually
be determined, and quantum coherence is not effectively lost.
An infinite value of $d_r$ is a necessary condition for an
effective loss of quantum coherence. We do not know the sufficient
conditions: we hope to return to this question in a future publication.

Finally, there is an important question we have been postponing:
why should there be superselection sectors among the black hole
groundstates? Of course they might arise as a consequence of
symmetries, but we have no strong reason to
believe that this is the case. Rather we are arguing that it is a
logical and interesting possibility. One piece of evidence is the
following: according to Hawking the black hole entropy is proportional to
the area and in particular is finite. This entropy should be identified as
$\ln d_r$, the entropy within a given superselection sector. On the other
hand the possibility of arbitrary values of global charges for black
holes suggests, as in \evan, that the number of groundstates is infinite.
This is consistent with finite $d_r$ only if there are indeed
superselection sectors. Thus apparently something
must give: either a) there are an infinite number of states within each
superselection sector, quantum coherence is effectively lost and
Hawking's area law for entropy is incorrect, b) the
superselection sectors have a finite number of states, the area law
is correct and quantum coherence is not truly lost, or c) that the black
hole does not behave like a quantum system.

\newsec{Concluding Comment}


In conclusion, we would like to comment on the the generality
of the previous discussion on quantum whiskers and incoherence.
It might appear that our discussion depended heavily on
specific properties of dilatonic black holes and did not
apply for example to Reissner-Nordstrom. At the beginning of the
previous section we assumed, partially motivated by \evan, that the
black hole is described by a quantum state, and that the only potential
source of information loss is tracing over that quantum state. More
generally, the whole system might be described by a density matrix
together with a $\Ssl$-matrix governing its evolution. However, these
matrices are subject to powerful physical
constraints such as probability conservation, locality and energy
conservation. It is plausible, though we have been unable to prove this,
that every system compatible with these constraints can be obtained by
tracing over some internal, unmeasurable quantum system modeling the
black hole (some results in this direction can be found in \BPS).
If this is true, then our considerations are quite
general and apply to scattering off of any type of extremal black hole.
If it is not true, it would be very interesting to characterize
counterexamples in which particle-hole
scattering can truly lead to information loss without
violating energy or probability conservation, or storing the information in
the black hole.
\vglue1cm
\centerline{\bf ACKNOWLEDGMENTS}
We wish to acknowledge the hospitality of the Aspen Center for Physics,
where part of this work was carried out.
We are grateful to M. Alford, C. Callan, J. Harvey and G. Horowitz for
useful conservations, and to John Preskill
for exciting our interest in the problem of particle-hole scattering
some time ago. As this manuscript was nearing completion, we received
a preprint \BDDL\ from Banks, Dabholkar, Douglas and O'Loughlin with
substantial overlap with some of the material in sections 5 and 6.
This work was supported in part by DOE grant DE-FG03-91ER40168 and by
an NSF PYI grant, PHY-9157463, to S.B.G.

\listrefs
\end